\documentclass[fleqn,10pt]{wlscirep}

\usepackage{amssymb,amsfonts,amsmath}
\usepackage{graphicx,color}
\usepackage[font=small,labelfont=bf,justification=justified]{caption}

\title{Preferential Root Tropism Induced by Structural Inhomogeneities in 2D Wet Granular Media }

\author[1,*,+]{Cesare M. Cejas}
\author[1]{Lawrence A. Hough}
\author[1]{Raphael Beaufret}
\author[1]{Jean-Christophe Castaing} 
\author[2]{Christian Fr\'etigny}
\author[1]{R\'emi Dreyfus}
\affil[1]{Complex Assemblies of Soft Matter, CNRS-Solvay-UPenn UMI 3254, Bristol, PA 19007-3624, USA}
\affil[2]{Sciences et Ing\'enierie de la Mati$\grave{e}$re Molle (SIMM) CNRS UMR 7615 ESPCI, 10 rue Vauquelin, Paris 75005 France}

\affil[*]{cesare.cejas@espci.fr}

\affil[+]{Microfluidics, MEMS, Nanostructures Laboratory, CNRS Chimie Biologie Innovation (CBI) UMR 8231, Institut Pierre Gilles de Gennes (IPGG), ESPCI Paris, PSL Research University, 6 rue Jean Calvin Paris 75005, France}

\keywords{roots, inhomogeneities, capillarity, granular media}

\begin{abstract}
\textbf{We investigate certain aspects of the physical mechanisms of root growth in a granular medium and how these roots adapt to changes in water distribution induced by the presence of structural inhomogeneities in the form of solid intrusions. Physical intrusions such as a square rod added into the 2D granular medium modify water distribution by maintaining robust capillary action, pumping water from the more saturated areas at the bottom of the cell towards the less saturated areas near the top of the cell while the rest of the medium is slowly devoid of water via evaporation. This water redistribution induces ``preferential tropism'' of roots, guiding the roots and permitting them to grow deeper into more saturated regions in the soil. This further allows more efficient access to available water in the deeper sections of the medium thereby resulting to increased plant lifetime.} 
\end{abstract}

\begin{document}
\flushbottom
\maketitle
\thispagestyle{empty}

%\noindent Please note: Abbreviations should be introduced at the first mention in the main text – no abbreviations lists. Suggested structure of main text (not enforced) is provided below.

%==========================================

%\section*{Introduction}
Root-water interactions remain an important topic in understanding transport to maintain robust water absorption in the face of today’s depleting resources. Such interactions are complex owing to the sensitivity of root systems and the presence of a wide variety of factors that affect not just large-scale (ecosystem level)~\cite{Pierret07, Hacke12} but also small-scale dynamics (sub-organismal or cellular)~\cite{Kramer95, Bengough11, Hacke12}. Fundamentally, research involving root systems are narrowed down in terms of physical, chemical, and/or biological variables. Root elongation is sensitive to external factors, e.g. light (phototropism) and gravity (gravitopism)~\cite{Ditengou08}. To quantify and qualify these responses, studies have used fluorescence to focus on molecular interactions that govern macroscopic root reflexes such as bending~\cite{Ditengou08}. In addition, roots are also sensitive to its environment. For instance, the granular nature of the soil can further complicate the growth process since the interactions between the grains and root systems could possibly increase the resistances~\cite{Abdalla59, Clark03}. For this reason, other studies have investigated the correlation between global root architecture and its interactions with the soil particles~\cite{Pierret07} and water uptake~\cite{Doussan06, Garrigues06, Cejas14b}. Literature on root mechanics has provided information on root interactions in various granular materials, often changing physical properties primarily for mechanical investigations such as compaction in terms of particle size~\cite{Abdalla69}, presence of intrusions~\cite{Cejas14} or effective strength of the growth substrate and its effect on root morphology~\cite{Silverberg12}. Root mechanics requires understanding of forces that the root exerts on the granular material. For this purpose, the choice of granular substrate is often photoelastic grains, where root contact with the grain triggers birefringence~\cite{Wendell12, Kolb12}. Roots have even served as inspiration for diggers due to their sensitive ability to respond to force inhomogeneities present in heterogeneous granular systems~\cite{Wendell11, Wendell12}. In addition to mechanical impedance, water stress also helps drive root elongation in granular medium~\cite{Bengough11}. Roots experience stress due to the minimal amount of water left in the medium~\cite{Whalley05a, Bengough11}, which can decrease cell division rates~\cite{Sacks97} and limit growth~\cite{Whalley05a, Bengough11}. Studies~\cite{Jaffe85, Takahashi91} have also shown that roots can sense a gradient of moisture in its environment and generally grow towards the region of higher moisture or saturation content.

Studying water extraction and root growth have developed due to increased interest in water retention~\cite{Garrigues06, Cejas14}, with the goal of improving water accessibility and availability for crop productivity~\cite{Hacke12}. To improve water retention, the most common method used nowadays is additives in the form of hydrogels. For this reason, some studies have focused on the dynamics of hydrogel swelling to trap water for longer periods of time~\cite{Wei13, Wei14}. Generally root growth rates are typically slower than water fluxes such as evaporation~\cite{Cejas14b} or even infiltration/drainage, and as a result, the root is immediately exposed to levels of water stress~\cite{Cejas14b}.  Thus, a gradient of water content saturation exists in the vertical direction, where greater quantity of water is found in the bottom (towards the groundwater table) than near the surface. This also induces a partially saturated zone~\cite{Cejas14b, Cejas17, Cejas18} or unsaturated zone\cite{Shokri08, Prat02, Yiotis12a} or vadose zone~\cite{Hunt08} containing hydraulic networks that link the deeper portions of the medium to the surface via capillary action.

What if we can mechanically induce capillary action, such that water from the deeper and more saturated areas is consistently pumped upwards towards the root zone? This change in water distribution can be generated by modifying the structure of the granular medium by introducing inhomogeneities~\cite{Cejas14}, whose purpose is to redistribute water from the bottom (an area of higher water saturation) to the top (an area of lower saturation). To our knowledge, this approach has never been elucidated and would provide interesting insights into the relationship between the root and the surrounding granular medium.

In this study, we aim to experimentally investigate the root growth behavior in response to a change in water distribution induced by the presence of structural inhomogeneities in the granular medium. Our choice of inhomogeneity is a solid intrusion that is inserted in the 2D granular medium. This solid intrusion extends throughout the depth of the growth cell, linking the upper portion of the medium near the surface to the fully wet zone at the bottom. We show in these experiments that root elongation can be influenced by the structural inhomogeneities inside the granular medium (``preferential tropism''), thereby inducing a change in water distribution that can significantly control direction and movement of root growth and the subsequent lifetime of a plant.

\begin{figure}[ht]
\centering
\includegraphics[width=5in]{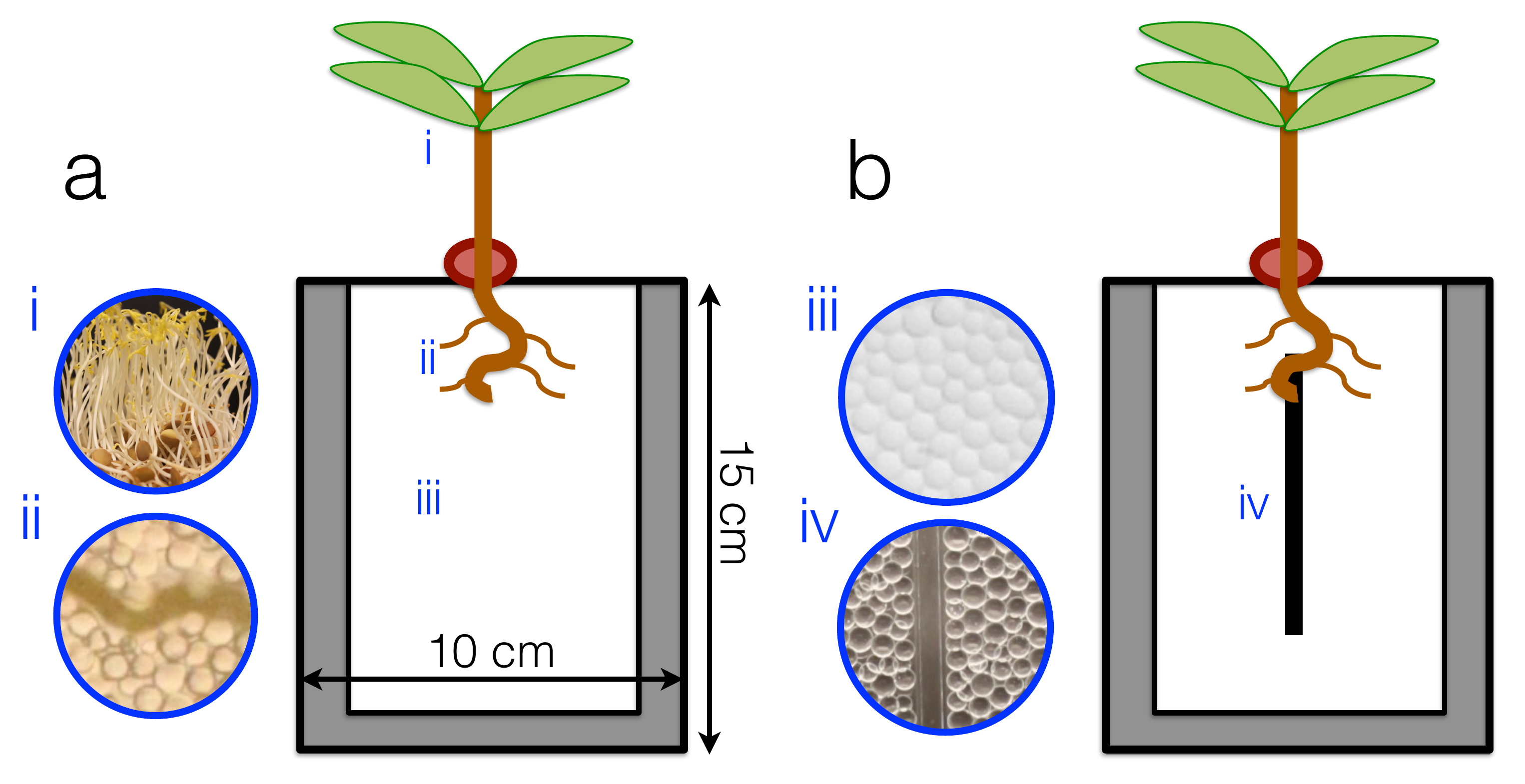}
\caption{Illustration of the Hele-Shaw cell used as a 2D root growth cell (rhizotron). Inset images show the different parts of the set-up. (\textbf{i}) lentil shoots, (\textbf{ii}) lentil roots within pore spaces between glass beads, (\textbf{iii}) glass beads (as model soil) fully saturated with water, (\textbf{iv}) image of the inhomogeneity or intrusion (square rod) inserted inside the 2D granular medium.}
 \label{Fig1}
\end{figure}

%-------------------------------------------------------------------------------------------------------------------
\section*{Results and Discussion}
\subsection*{Root Morphology Characterization}

Roots are extremely sensitive to environmental conditions~\cite{Desmet12} as well as changes in water distribution in the medium~\cite{Doussan06}, especially if it leads to water stress induced by drainage or evaporation. Hydraulic (or water) stress is defined as a condition where water content is reduced or depleted~\cite{Bengough11}. Roots are grown in different types of media, such as agar~\cite{Dubrovsky12}, sand~\cite{Mia96}, deformable gels~\cite{Silverberg12}, photoelastic grains~\cite{Wendell12}, or glass beads~\cite{Futsaether02}. The type of granular material is important since soil structure affects root growth~\cite{Passioura91}. Regardless of the choice of the growth substrate, it is imperative to characterize root growth in the chosen substrate to first understand how such roots adapt to their idealized environment and to use this information as a reference when comparing root growth subject to different physico-chemical treatments.
 
Root growth studies are commonly performed through monitoring a series of changes in root elongation as a function of time. However, they can be challenging because roots naturally grow in opaque or subsurface systems~\cite{Desmet12}. While 3D imaging systems such as X-ray tomography or neutron tomography have been gaining traction in the community~\cite{Robinson08, Weng18}, there are generally hardly accessible and are almost always expensive. Thus, root investigations have often relied on simple set-ups, normally in two-dimensions (2D) as shown in Fig.1. These techniques called rhizotrons~\cite{Futsaether02, Garrigues06} permit non-invasive spatial assessement of the distribution of the root as it develops in the granular medium. The method, when back-illuminated with a light box, also potentially permits simultaneous observation of water content over time. For this reason, the choice of granular material is often transparent.

We characterize root growth in such rhizotron systems, where we let the roots grow under a controlled environment (see Materials and Methods for details). We use lentils ($\it{Lens~culinaris}$) because lentil growth occurs at relatively rapid rates and they develop relatively simpler root systems - meaning lentils possess a distinct large primary root and a few observable secondary roots. Lentils have also been previously grown in 2D chambers~\cite{Futsaether02} and results show that they still exhibit robust growth and that their growth in 2D model systems are still comparable to that of real soil systems in 3D~\cite{Mia96}. 

\begin{figure}[ht]
\centering
\includegraphics[width=7in]{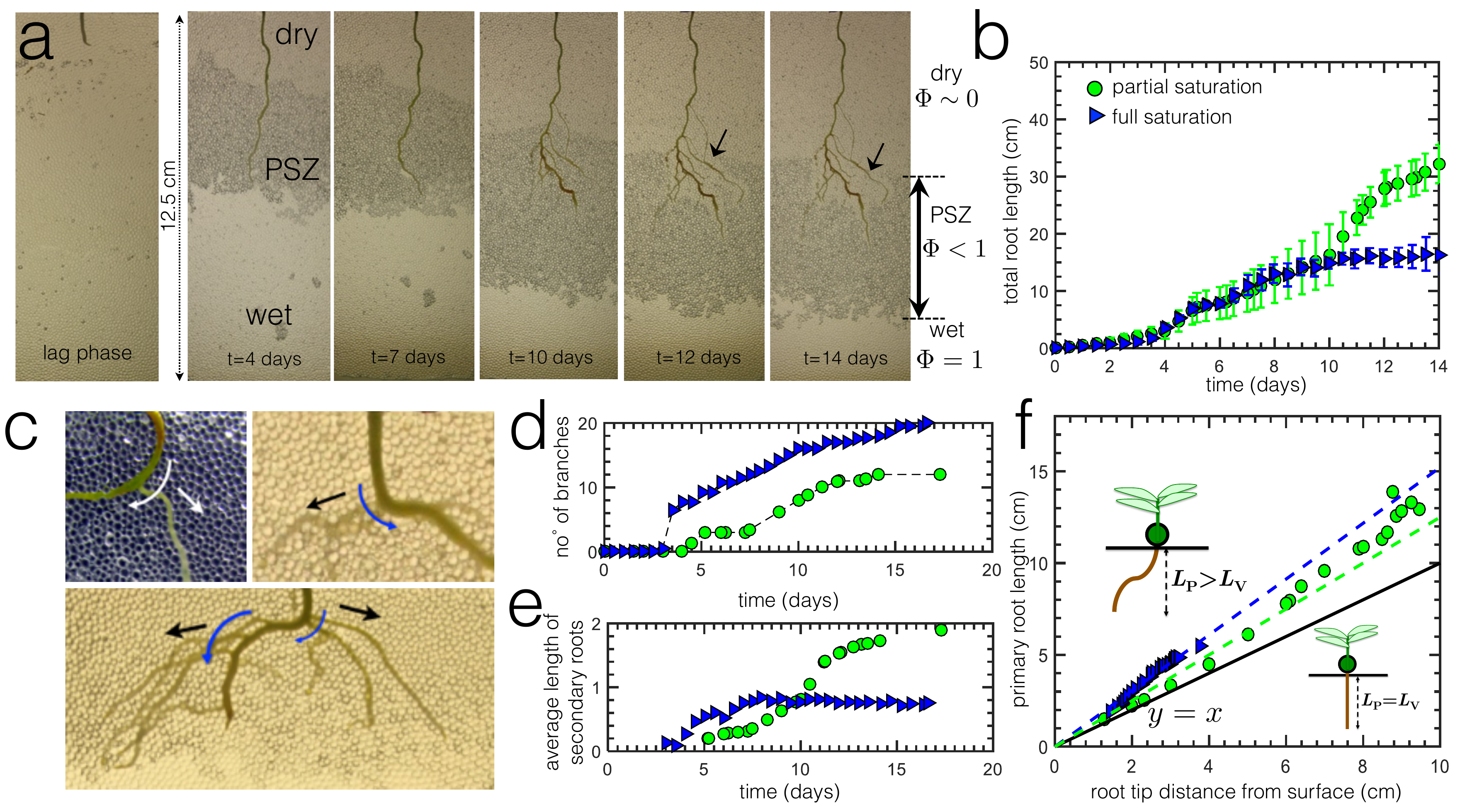}
\caption{(\textbf{a}) Typical images of root growth under partial saturation conditions, where evaporation is allowed to take place after the initial lag phase. We observe three distinct zones (dry, partially saturated, wet). (\textbf{b}) Total root length as function of time for both partial saturation and full saturation conditions after the initial lag phase. (\textbf{c}) Images showing emergence of lateral roots from convex sides (outer curves). Curved arrows signify outer curvature while straight arrows indicate growth direction. (\textbf{d}) Number of branching points as a function of time. (\textbf{e}) Average length of secondary roots normalized by the number of branching points. (\textbf{f}) Primary root length as function of root tip distance from suface. The solid line represents $y=x$, which corresponds to the case when the root grows in a straight vertical manner, perpendicular to the surface.}
 \label{Fig2}
\end{figure}

The roots are grown and all experiments are performed under the same conditions but repeated at different times, showing reproducibility of the root system in this particular 2D experimental set-up. Once a radicle germinates rom the seed, it is transplanted on top of the Hele-Shaw cell. At the start for all experiments, the 2D cells are filled with water up to the brim and thus overall water saturation, $\Phi$, in the medium is $\Phi\sim1$. It is expected that after germination, root growth is normally slow because the roots are still developing the necessary biological functions dedicated tor root functions. This initial slow growth phase is termed as lag phase, where radicles are most vulnerable~\cite{Fisher96}. For this reason, after germination, the roots are grown in partial saturation conditions for about 2 days maximum. This implies that water, mixed with nutrient solution, is constantly replenished into the Hele-Shaw cell at regular intervals such that the young radicle does not experience sudden water stress via evaporation.  As water evaporates, the granular system is slowly being depleted with water such that $0<\Phi<1$. This leads to the development of a partially saturated zone (PSZ), a mixture of air and water, ~\cite{Cejas14b, Yiotis12a, Lehmann08, Shokri08, Prat02,  Cejas17, Cejas18} appearing inside the granular medium. The bottom of the cell remains fully saturated. After 2 days, the water in the medium is allowed to freely evaporate without additional replenishment. 

We characterize the root morphologies in partial saturation conditions by also comparing it to root growth in full saturation conditions under controlled ambient atmospheric conditions ($T=23^{\circ}\pm2$, $R_H=45.0\pm5.0\%$). In the latter condition, water (mixed with nutrient solution) is constantly being replenished at regular intervals even after the initial lag phase such that the root, at any point during the experiment, does not experience immediate severe water stress. While in the former condition, as explained in the previous paragraph, evaporation is permitted after the lag phase. This eventually results to the formation of PSZ, which induces hydraulic stress to the roots~\cite{Bengough11}. A typical example, Fig. 2a, shows the temporal evolution of root growth as well as the PSZ evolution during evaporation in the presence of such root systems. During evaporation, the PSZ initially develops around the root zone but gradually grows further inside the medium, distancing itself away from the root zone and a dry zone eventually appears just below the cell surface. Over time, the dry zone eventually catches up and a fraction of the entire root system is exposed to this dry zone. Root activity within the PSZ can be manifested by depletion zones in the vicinity of the root, as reported in previous studies~\cite{Cejas14b}. Since evaporation rates are typically faster than root growth rates~\cite{Cejas14b, Cejas13}, the root is then completely engulfed by the developing dry region. At this point, the root eventually dies without being able to take advantage of the considerable quantity of water left unused at the bottom of the cell, far from the root zone. 

In Fig. 2b, we compare the root growth in both full and partial saturation conditions, where we measure the total root length (primary + secondary). Results reveal that overall root systems grow $2\times$ more in partially saturated systems, where a mixture of air and water exists. The onset of hydraulic stress results to roots developing new structures~\cite{Passioura91}. The graph of total root length follows a sigmoidal pattern~\cite{Fisher96}, with a plateau towards the end. In Fig. 2c, we show some examples of lateral roots emerging from the convex sides of the root. Literature points out that this seems to be related to the auxin plant hormone distribution~\cite{Laskowski08} because high concentration of auxins at the curvature of roots triggers lateral organ formation of branching points for the secondary lateral roots~\cite{Ditengou08}. We also measure the average length of the secondary roots, normalized by the number of branching points emerging from the primary root. We observe that, for this root sample, tertiary roots (roots that emerge from secondary branches) do not form as much within the experimental observation window. Thus, the total root length mainly comprises of the primary and the secondary lateral roots. The graph in Fig. 2d shows that more branches grow in fully saturated conditions but Fig. 2e shows that despite the greater branching points, the resulting secondary roots under full saturation conditions ultimately have smaller lengths than in partial saturation conditions. Roots elongate in partially saturated systems much further in search of water. Finally, in Fig. 2f, we measure primary root length, $L_P$, with respect to the vertical distance between primary root tip and the surface, $L_v$. The solid line represents $y = x$, which corresponds to the case when the root grows in a straight vertical manner perpendicular to the surface. Due to the presence of granular particles, root growth is never straight so any deviation from this solid line can be seen as a measure of the root's tortuosity. Roots in partially saturated systems grow deeper and deviate less from the line suggesting the necessity of roots to find areas of higher saturation using the shortest possible distance. 
%------------------------------------------------------------------------------------------------------------------------------------
\subsection*{Strategic Modification of Water Distribution: Capillary Action}

\begin{figure}[ht]
\centering
\includegraphics[width=7in]{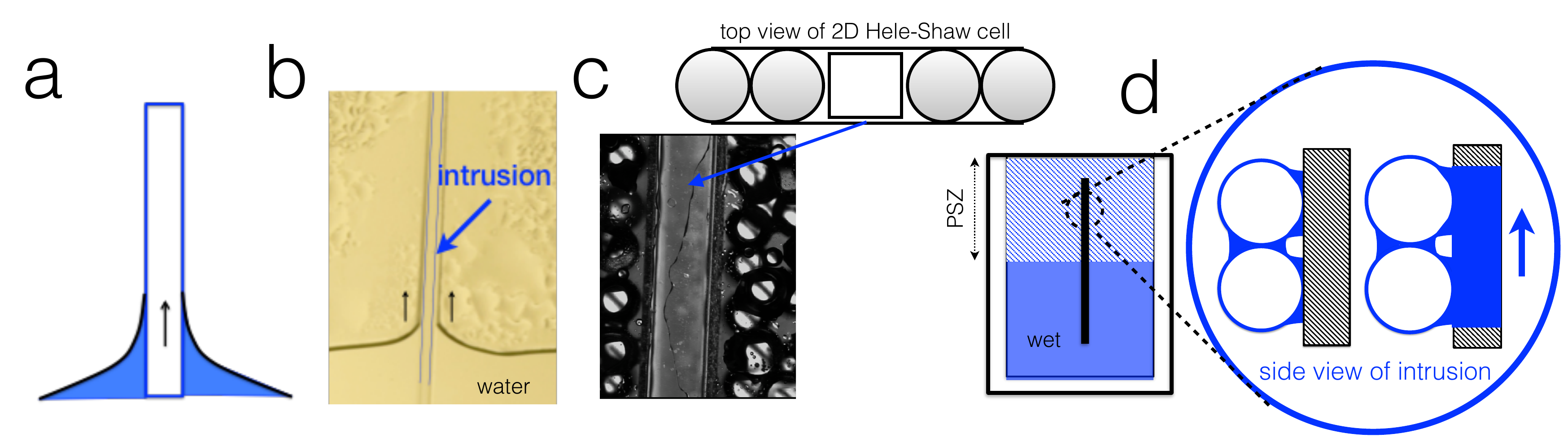}
\caption{ (\textbf{a}) Illustration of the continued capillary rise in the vicinity of the intrusion when inserted in the 2D Hele-Shaw cell filled only with water.  (\textbf{b}) Experimental images showing water flow along the length of the inserted intrusion. (\textbf{c}) Magnified image of the intrusion wall in contact with the transparent glass wall showing a liquid film. (\textbf{d}) Illustration of the PSZ with the intrusion. A magnified image further illustrates how a coalescence of liquid films between beads and the wall could possible create a flowing liquid film. }
 \label{Fig3}
\end{figure}

From experiments, it is evident that the roots are never able to grow sufficienty fast enough to reach the fully wet region at the bottom of the cell compared to the rate of water loss from evaporation and transpiration. It is therefore imperative to find solutions to redistribute water from the more saturated area at the bottom of the cell back to the root zone. One solution is to induce a robust capillary action. Hydraulic networks in the PSZ, formed from liquid film connections,  already drive water upward via capillarity~\cite{Cejas14b, Yiotis12a, Lehmann08, Shokri08, Prat02,  Cejas17, Cejas18} but these networks eventually disconnect as liquid films thin out from evaporation. Thus, to replenish these films, we must constantly induce capillarity even as the surrounding water continues to evaporate. This can be achieved by inserting a piece of solid intrusion in the form of a square rod inside the 2D medium. This not only structurally modifies the model soil but can also induce a gradient of water distribution through wetting. Thin rods surrounded with water induce liquid to rise in the vicinity of the rod due to capillary action~\cite{deGennes10}, schematically shown in Fig. 3a and experimentally in Fig. 3b. As water evaporates, capillary action drives water from the deep (and more saturated) parts of the soil to the upper part of the soil to replenish the quantity of water lost from evaporation, thereby modifying water distribution in the granular medium in a way that shifts water content closer to the root zone~\cite{Cejas14b}.  

In this 2D set-up, this rod has approximately the same thickness of the 2D cell, where the insertion of the square rod generates an area of contact between the rod wall and the wall of the 2D Hele-Shaw cell. This space has a characteristically small size, Fig. 3c, that allows a thin liquid film to be robustly pumped upward via capillary action. As the PSZ continues to develop due to evaporation further inside the medium, water is also simultaneously being pumped near the intrusion. In addition, especially in the presence of granular particles, the coalescence of the liquid bridges between the intrusion and the nearby glass beads aids the hydraulic pathway as schematically shown in Fig. 3d. This water redistribution is an effect of the structural modification of the granular medium due to an inhomogeneity in the form of a solid intrusion~\cite{Cejas14, Cejas13}. 

%-----------------------------------------------------------------------------------------------------------------------------------------
\subsection*{Preferential Tropism and Changes in Root Morphology}

\begin{figure}[ht]
\centering
\includegraphics[width=7in]{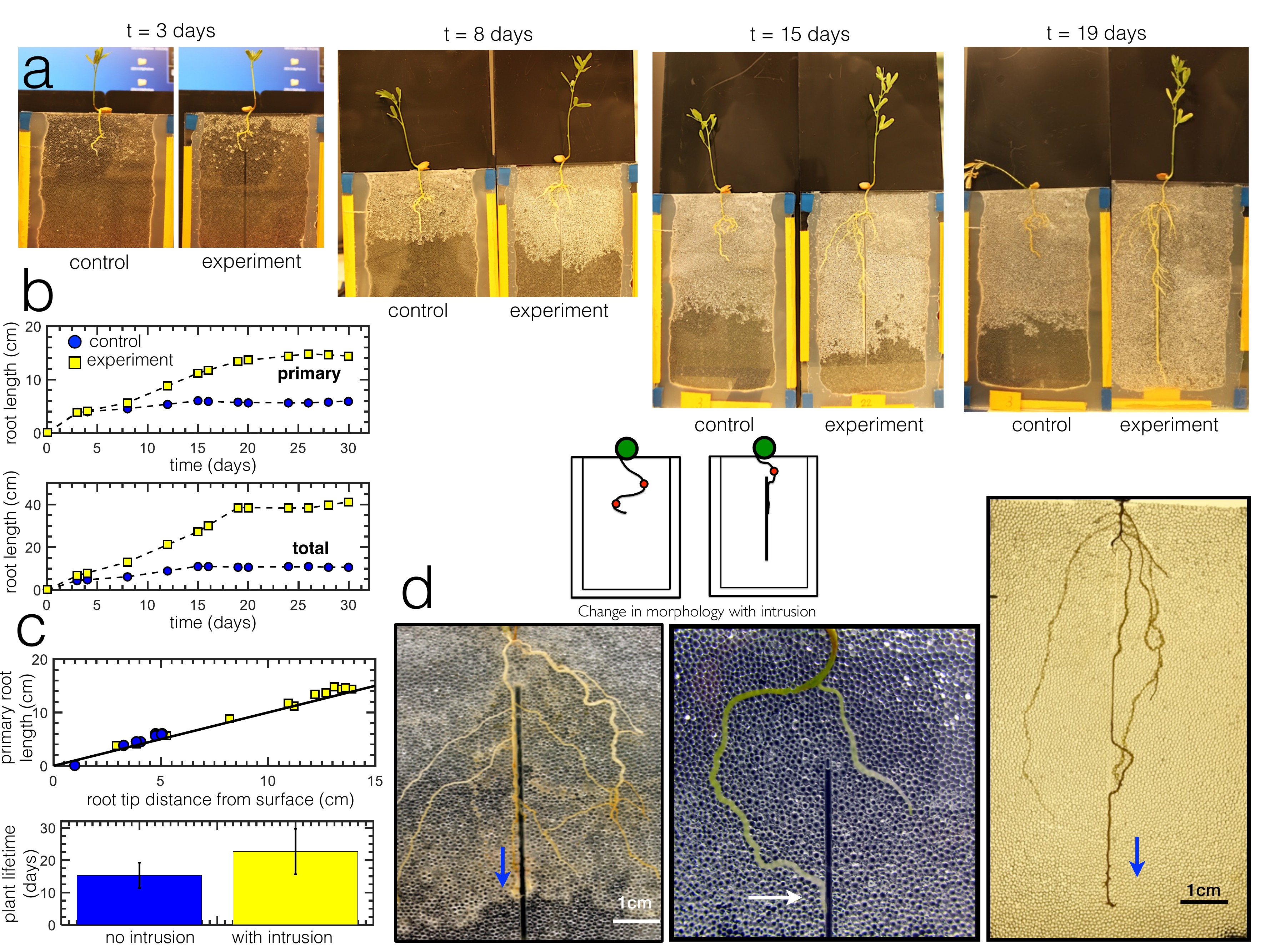}
\caption{(\textbf{a}) Typical example of time evolution of root growth experiment with intrusion (experimental) and without intrusion (control). Both experiment and control set-ups are performed simultaneously under partial saturation conditions. Results show a change in root morphology and a distinct vertical growth pattern in the presence of the intrusion. (\textbf{b}) (top) Primary root growth curve for the images in (a). (bottom) Total root growth curve (primary + secondary) for the images in (a).  Results show root growth is greater in the presence of the intrusion.  (\textbf{c}) (top) Primary root length as function of root tip distance from suface. The solid line represents $y=x$, which corresponds to the case when the root grows in a straight vertical manner, perpendicular to the surface. In the presence of the intrusion, the root follows a vertical growth pattern. (bottom) Bar graph showing average lifetimes of all experiments performed with and without intrusion. (\textbf{d}) Illustration of the observed differences in root morphology with and without intrusion as well as some additional experimental images showing preferential tropism beside the intrusion. In some cases, the root initially deviates away but later finds the intrusion. }
 \label{Fig4}
\end{figure}

Then, we couple the structural modification of the granular media to actual root growth experiments. In each of the experimental runs, two roots are grown simultaneously and always under the same conditions: a control cell with roots with no intrusion and an experimental cell with both roots and the intrusion. Both systems are investigated under partially saturated conditions, meaning water loss via evaporation is allowed to take it course after the initial lag phase. The cells are briefly taken out of the controlled environment chamber during image acquisition at different time intervals. During days when we are unable to take pictures, we perform interpolation of the root length between measured experimental values.

We show in Fig.4a typical experimental images of root growth with and without intrusion, under the same growth conditions. We grow a control experiment (without intrusion) simultaneously with the variable experiment (with intrusion) under the same conditions to easily compare the elongation behavior of the two systems at a given time. When another experimental run is performed, both control and variable systems are always prepared. This way, with all the conditions being the same, the effect of the physical intrusion could be readily differentiated.  We perform a total of 11 experimental runs, where each run consists of a pair or control (without intrusion) and experimental cell (with intrusion), where root growth is performed and analyzed at the same time and under the same conditions.

In comparing control and experimental set-ups, the final root branching architecture is also another aspect to consider~\cite{Doussan06, Dupuy05}.  Experimental results in Fig.4a strongly show distinct root morphology in the presence of the solid intrusion. This is evidenced from the vertical growth pattern of the primary root that is typically absent in a standard 2D granular medium alone (without inrusion). The vertical growth pattern occurs adjacent to the intrusion and that there is preferential tropism in response to the water distribution that results from evaporation and the presence of the intrusion.  As the PSZ recedes due to evaporation towards deeper sections of the cell, the effect of the intrusion also guides the roots deeper and allows them to stay within a PSZ, thus also allowing the roots to proliferate further.  Here, we define ``preferential tropism'' as a general type of root response that favors a certain growth direction due to what it perceives as changes in its surroundings. In this case, the root response is triggered by the robust capillary action in the vicinity of the intrusion.

General analysis of root elongation, typically such as in Fig. 4b (top), shows that the primary root grows much longer in the presence of the intrusion, guiding the root towards the deeper and more saturated areas of the medium - something that is difficult for the roots to achieve without the intrusion. In Fig. 4b (bottom), the total root growth also is in favor of the presence of the intrusion. The graph in Fig. 4c (top) shows that the growth of the primary root is nearly vertical and straight, almost parallel to the inhomogeneity. Because the roots in the presence of the intrusion has greater access to water in response to the stress induced by evaporation, they generally thrive longer as shown in the comparison of plant lifetimes in Fig. 4c (bottom). Plant lifetime was determined from experimental images, where a qualitative deterioration of plant shape and form (dehydration, wilting, etc) suggests it is already dead. Measurements of its lifetime (duration of plant life before death) in both experiments with and without intrusion show that the presence of the intrusion increases overall plant lifetime by approximately 1.5-2 times.

Some additional images  further show the robustness and reproducibility of the effect of the physical intrusion on total root growth. In some experiments, roots initially deviate away from the intrusion, Fig. 4d, probably due to physical limitations (e.g. unable to sufficiently push granular particles aside). However, the roots later find the intrusion. Repeated experimental results show that the roots consistently ``sense'' this robust capillary action near the intrusion while the rest of the medium around it is being de-saturated with water from evaporation. As a result, the roots elongate in the direction of the intrusion. Eventually, the roots also grow adjacent to it.

Root elongation can differ even for the same root systems, {\it{i.e.}} its total macroscopic root morphology changes depending on how roots interact with the soil. Experimental results, however, strongly shows a distinct root morphology in the presence of the solid intrusion. This is evidenced from the vertical growth pattern of the primary root that is normally absent without the intrusion. The vertical growth pattern occurs adjacent to the intrusion.

%--------------------------------------------------------------------------------------------------------------------------------
\subsection*{Root Behavior Induced by Inhomogeneities}

\begin{figure}[ht]
\centering
\includegraphics[width=7in]{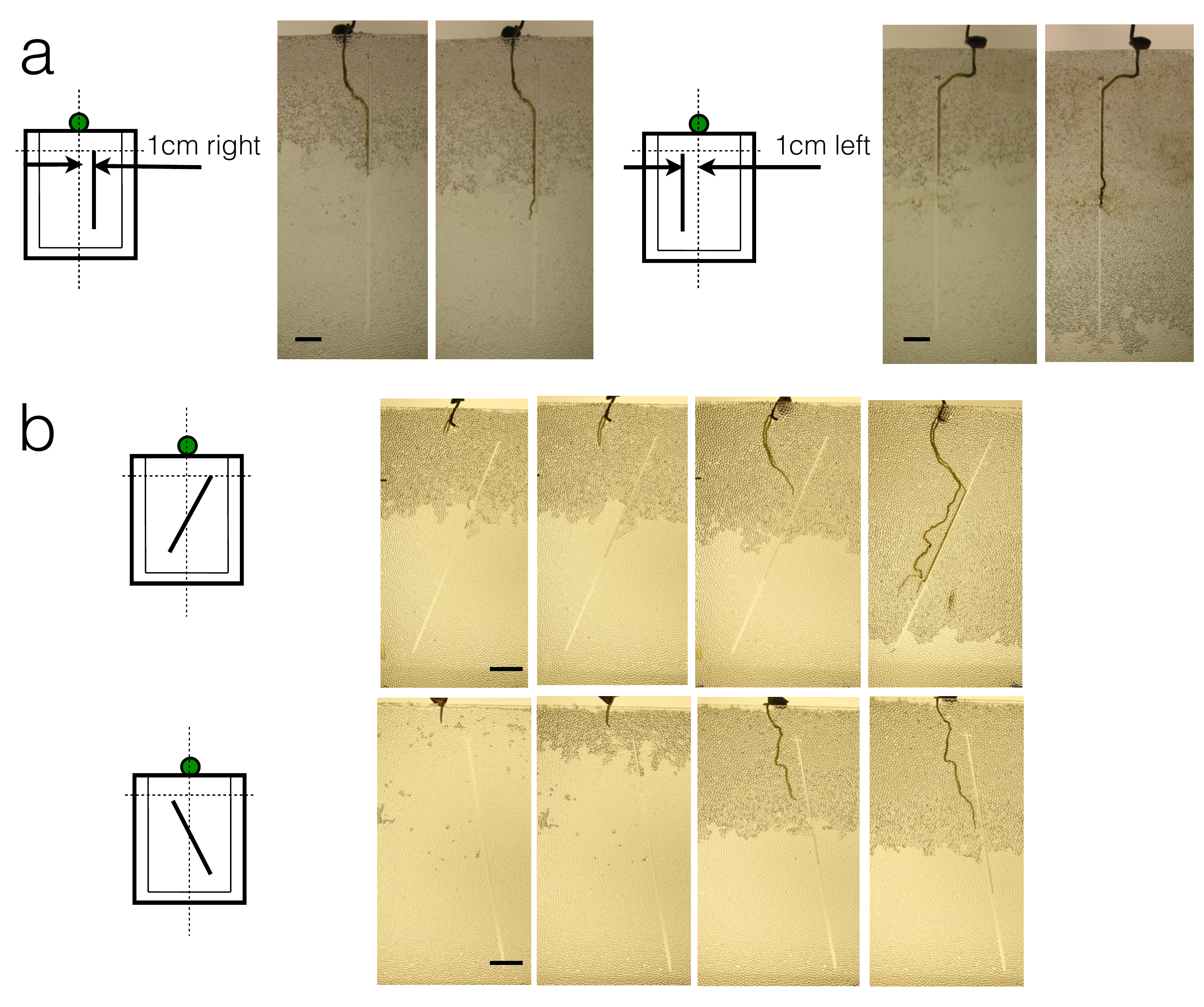}
\caption{ Experimental images of root growth showing preferential tropism of intrusion at varying distances and orientations. (\textbf{a}) Intrusions are oriented vertically but 1cm both left and right from the center. (\textbf{b}) Intrusions are oriented diagonally. In all cases, scale bar is 1cm. }
 \label{Fig5}
\end{figure}

The solid rod intrusions modify water distribution in the granular medium and this has consistently demonstrated preferential tropism. Thus far, the seed has relatively easy access to the intrusions due to the intrusion being positioned directly beneath it. Sometimes, however, as seen in an image in Fig.4d, the roots can initially deviate before finding the intrusion. This raises the question of characteristic distance: how far must the seed be from the intrusion to still be able to sense the intrusion? To offer insights into the question, we perform experiments by deliberately varying the distance of the intrusion about 1cm away on either side from the initial root as shown in Fig. 5a. Results consistently indicate that the root  continues to develop in the direction of the intrusion, assisted principally by the growth of its secondary roots. We have also performed a modified version of the problem by putting the intrusion in an oblique position also shown in Fig. 5b. In this manner, we demonstrate competition between capillarity along the intrusion wall and gravity. Results once again show consistency of preferential tropism towards the intrusion despite the less favorable position of the intrusion. This strongly suggests that while roots generally respond to gravity, they prefer to grow towards areas of higher saturation~\cite{Whalley05a}.

Root penetration in soils is limited to a number of soil mechanical properties~\cite{Bengough11}, such as particle size. Smaller particles provide mechanical impedance and can reduce root proliferation~\cite{Clark03}. Since soil characteristics challenge root elongation~\cite{Whalley05}, it has been shown in studies that roots tend to cluster in macropores or areas of low resistances~\cite{Whalley05a}. Nevertheless, we perform additional experiments to confirm the result that the phenomenon of preferential tropism is in response to capillary action and not as a result of lesser granular packing. 

The robust root response to the physical intrusion indeed raises additional interesting questions on the mechanism of preferential tropism. One may argue that perhaps the difference in local granular packing adjacent to the intrusion and the rest of medium might be responsible for the preferential tropism, due to decreased mechanical impedance in this vicinity. The packing beside the intrusion has 30 percent more void space, and is thus the perfect candidate for the path of least resistance. However, additional experiments, shown in Fig. 6, performed in the presence of the intrusion but under conditions of full saturation, {\it{i.e.}} water is constantly replenished throughout the experiment, have shown that roots do not find the intrusion. Under full saturation where evaporation is not allowed, Fig. 6a demonstrates that root growth (both primary and total) in the presence of the intrusion is comparable to that without the intrusion. As a result, the intrusion under full saturation does not induce a significant effect and root morphologies are similar, even the average length of the secondary roots as shown in Fig. 6b. The inset in Fig. 6b reveals an experimental image taken at $t=14$~days, showing that under full saturation the root does not sense the intrusion, i.e. no preferential tropism exists. Furthermore, in Fig. 6c, the deviation of the primary root from the vertical line $y=x$ is also similar for both experiments, thereby also suggesting similar root morphologies under full saturation conditions. Thus, the root are unable to find the intrusion under conditions of full saturation since water is always readily available.

In contrast to partial saturation conditions, water content decreases elsewhere in the PSZ via evaporation and thus the liquid film along the intrusion wall flows robustly by capillary action. The contact of the solid intrusion with the wall of the growth cell leaves a tiny area that generates enough pressure to drive capillary action from the bottom to the top. The flow is also aided by a coalescence of liquid bridges between adjacent beads and the intrusion. These original experimental results suggest that the root senses the capillary action along the intrusion wall as the percolation front deepens and the PSZ develops.

\begin{figure}[ht]
\centering
\includegraphics[width=7in]{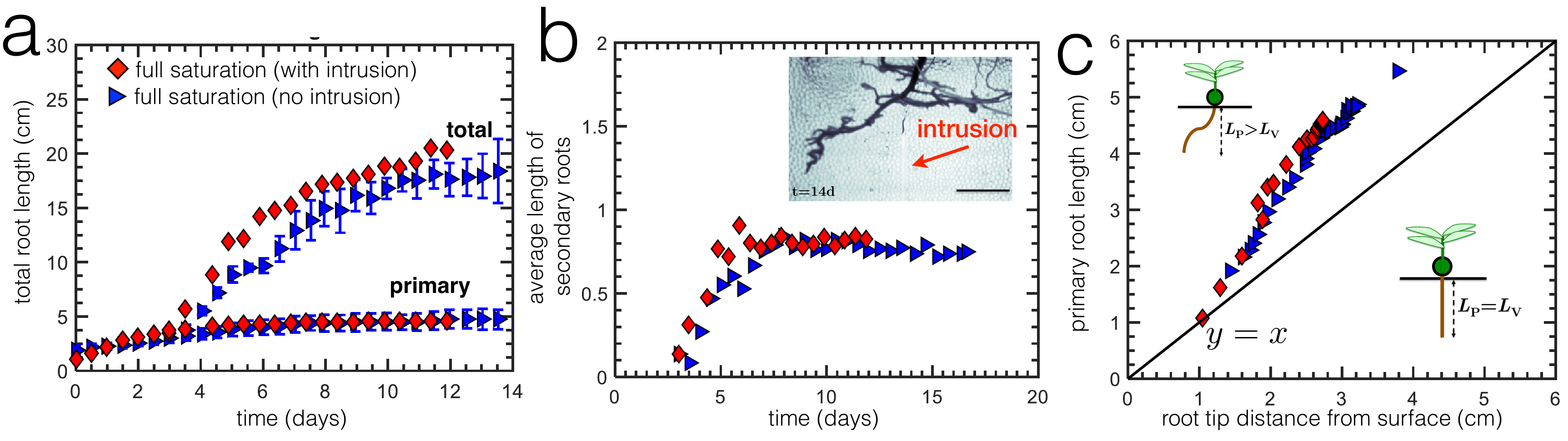}
\caption{Comparison of root growth with and without intrusion under conditions of full saturation. (\textbf{a}) Primary and total root length. (\textbf{b}) Average number of secondary roots normalized by the nu,ber of branching points. The inset figure shows an experimental image at $t=14$~days, where the root still does not find the intrusion under full saturation. The scale bar is 1cm. (\textbf{c}) Root length as a function of tip distance. The solid line is $y = x$ suggests that the root grows in a straight vertical manner perpendicular to the surface. Results show that under full saturation, the intrusion does not significantly change the root elongation and root morphology.}
 \label{Fig6}
\end{figure}

%----------------------------------------------------------------------------------------------------------------------
\section*{Conclusion}
Inhomogeneities in granular media have shown to actively and physically influence root elongation caused by the redistribution of water content in the medium. The capillary action along the intrusion is a primary result of the 2D nature of the set-up and depends on the small contact area between the intrusion and cell wall. From a fundamental perspective, it is interesting to understand how a change in the size of this contact area affects preferential tropism and, in addition, how such 2D effect can be translated into real 3D systems. On the other hand, from an applications perspective, root growth is important for increasing plant lifetime, which also improves agricultural yield. Future studies must focus on how root growth adapts itself with respect to different levels of water saturation in the medium. This serves as a jump-off point for designing better additives that not just store water but also guide roots to more saturated regions inside the medium; therefore maximizing water usage and improving efforts to increases plant lifetime - key aspects in applications involving water retention.

%-------------------------------------------------------------------------------------------------------------------------------------------
%\subsection*{Subsection}
%Example text under a subsection. Bulleted lists may be used where appropriate, e.g.
%
%\begin{itemize}
%\item First item
%\item Second item
%\end{itemize}
%
%\subsubsection*{Third-level section}
% 
%Topical subheadings are allowed.
%
%\section*{Discussion}
%
%The Discussion should be succinct and must not contain subheadings.
%-----------------------------------------------------------------------------------------------------------------------
\section*{Materials and Methods}
\textbf{2D growth chambers.} The root growth cell is a Hele-Shaw cell built from two glass sheets of size 10~cm (length) and 15~cm (width). All the sides, except the upper part, are sealed with commercial silicon paste. Due to the presence of the paste, the effective dimension of the porous medium is reduced to about $7.5$~cm$\times$$13$~cm. To ensure that the cells contain a monolayer stack of glass beads, $d = 1 \pm 0.2$mm (borosilicate, Sigma-Aldrich, USA), we a put a spacer with a thickness approximately equivalent to the diameter of the glass beads. The glass beads initially have around 20$\%$ polydispersity, but are sieved prior to the experiment to only use monodisperse sizes as possible. Despite the sieving process, the Hele-Shaw cell is still quasi-2D since a perfect 2D system cannot be achieved. We then wash the glass sheets with hydrophobic silane solution (OMS Chemicals, Canada) for two hours to reduce wetting effects along the glass wall. The glass beads are all hydrophilic for all root growth experiments, washed with $0.1$ M HCl and dried in an oven overnight at $70^{\circ}$C. Contact angle values are $82 \pm 4^{\circ}$ for hydrophobic glass sheet and $16 \pm 4^{\circ}$ for hydrophilic glass beads. 

\textbf{Root growth set-up.} An illustration of the experimental set-up is presented in Fig.1. We use lentils ($Lens$~$culinaris$) as the model root because they grow relatively fast and develop a distinct primary root with observable secondary roots.

The growth conditions are kept controlled as much as possible. The roots are grown in an air-conditioned laboratory, inside growth chambers, under ambient relative humidity values, $H_R = 45\pm5\%$ and at ambient temperature, $T = 23\pm2^{\circ}$C. The lentil seeds are first allowed to germinate in moist paper or tissue in the dark for 2-3 days. Once a radicle has developed from the seed, it is then placed on top of the Hele-Shaw cell. The initial condition of the cell is such that its water saturation, defined as $\Phi$, is full. This means that $\Phi = 1$ at $t = 0$, where $t$ is time. We use a vacuum water pump to initially saturate the medium by immersing the cell in the water pump filled with the liquid. This apparatus simply removes air from the medium, allowing the surrounding water to percolate and enter the medium. The average length of the radicle that has sprouted at the time of seed transplant onto the top surface of the Hele-Shaw cell is $L_T = 1.5\pm0.5$~cm. It is expected that the initial phase of root growth would be normally slow~\cite{Fisher96} since roots are still developing the necessary biological functions dedicated to certain root processes. This is the initial lag phase~\cite{Fisher96}. For this reason, for the first few days (up to $\sim2$ days after transplant), water loss is being regularly replenished. Under conditions of full saturation, water loss from evaporation is constantly being replenished while under conditions of partial saturation, evaporation is allowed to take its course after the initial lag phase. The water used actually contains a small amount of Hoagland solution (1/4 volume ratio) to provide the necessary nutrients for plant growth. Nutrients are especially crucial during the developing phase of the seeds. The plants were constantly grown under grow lights with light intensity of $\sim 40\mu$mol/m$^{2}$s, which is comparable to white lamps used in literature~\cite{Futsaether02}.

\textbf{Image acquisition.} We take images of root growth as function of time generally using Canon 500D SLR camera at certain intervals since root elongation time scales are normally slow. We use $18-55$~mm lens and the spatial resolution was about $35\mu$m per pixel. We use a light box to back-illuminate the 2D cell to completely distinguish the main components: roots, beads, and water. Water content can be estimated from intensity of a transmitted light~\cite{Doussan06}. However, the resolution does not permit accurate assessment of water saturation using this technique. 

The obvious challenge when studying root systems is that the duration of the experiment usually lasts relatively long. For lentils, growth characterized by a fully developed root and shoot system is observed within 12-18 days. In addition, experimental investigations on root systems need to be performed several times to ensure that an observed phenomenon is indeed robust and reproducible. The slow time scales involved with studying root growth normally requires simultaneous experiments to obtain as much results as possible within a limited time frame. The lack of multiple Canon 500D SLRs required use of commercial webcams as supplementary imaging tools. These webcams have been automated using a Matlab program.

We measure root length via the segmentation method. The methods breaks down a continuous curve into smaller line segments, whose individual lengths have been calculated from a pixel-to-length calibration that is performed before each measurement. Root length measurement via image substraction is often difficult to perform with these experiments because sometimes the thin secondary lateral roots in the PSZ are too faint. Roots in granular systems normally adopt a curved profile. As a result, the total contour length is measured by creating smaller line segments. Preliminary tests of the segmentation method performed on various sets of curves of known lengths reveal a measurement error of $\pm5\%$. The total length is measured from the summation of the lengths of the individual line segments.

%---------------------------------------------------------------------------------------------------------------------------------------
\bibliography{Roots_References}

%---------------------------------------------------------------------------------------------------------------------------------------
\section*{Acknowledgements}

We thank the support of UMI 3254 joint laboratory of CNRS, Solvay, and the University of Pennsylvania (UPenn). We also thank Prof. Doug Durian (UPenn), Zhiyun Chen (Solvay), and Ryan Murphy (Solvay) for engaging discussions.
%---------------------------------------------------------------------------------------------------------------------------------------
\section*{Author contributions statement}

C.M.C performed experiments, image analysis, and treatment;  R.B. helped perform experiments, L.H. and J.C.C. provided technical direction from an applications perspective while C.F. and R.D. provided scientific direction from a fundamental perspective; C.M.C. and R.D. wrote the manuscript. 

%---------------------------------------------------------------------------------------------------------------------------------------
\section*{Data availability statement}

The datasets generated during and/or analyzed during this study are available from the corresponding author on reasonable request.

\section*{Competing financial interests}

The authors declare no competing financial interests. 

%The corresponding author is responsible for submitting a \href{http://www.nature.com/srep/policies/index.html#competing}{competing financial interests statement} on behalf of all authors of the paper. This statement must be included in the submitted article file.

%\begin{table}[ht]
%\centering
%\begin{tabular}{|l|l|l|}
%\hline
%Condition & n & p \\
%\hline
%A & 5 & 0.1 \\
%\hline
%B & 10 & 0.01 \\
%\hline
%\end{tabular}
%\caption{\label{tab:example}Legend (350 words max). Example legend text.}
%\end{table}
%
%Figures and tables can be referenced in LaTeX using the ref command, e.g. Figure \ref{fig:stream} and Table \ref{tab:example}.

\end{document}